\documentclass[conference]{IEEEtran}
\usepackage[noadjust]{cite}  

\usepackage{graphicx}
\usepackage[normalem]{ulem}
\usepackage{fancyhdr}
\usepackage[labelfont=bf, textfont=bf, skip=2pt]{caption}
\usepackage[english]{babel}
\usepackage{booktabs}
\usepackage{multirow}
\usepackage{url}
\usepackage{pifont}
\usepackage{footnote}
\usepackage{tikz}
\usepackage{enumitem}
\usepackage{ltablex}
\usepackage{xargs} 
\usepackage[first=0,last=9]{lcg}
\usepackage{color, colortbl}
\usepackage[colorinlistoftodos,prependcaption,textsize=tiny]{todonotes}
\usepackage{marginnote}
\usepackage{clipboard}
\usepackage{subfig}
\usepackage{mathtools}
\usepackage[framemethod=tikz]{mdframed}
\usepackage{setspace}
\usepackage{blindtext}
\usepackage{imakeidx}
\usepackage{soul}
\usepackage{hyperref}
\usepackage{blindtext}
\usepackage{imakeidx}
\usepackage{graphbox}
\usepackage{dblfloatfix}

\usepackage{balance}

\newcommand\ignore[1]{ }

\definecolor{dred}{rgb}{0.50, 0.00, 0.00}
\definecolor{dora}{rgb}{0.80, 0.60, 0.00}
\definecolor{dpur}{rgb}{0.60, 0.00, 0.60}
\definecolor{dgre}{rgb}{0.00, 0.80, 0.80}
\definecolor{dblu}{rgb}{0.00, 0.00, 0.80}
\newcommand{\om}[1]{\textcolor{black}{#1}}

\usepackage[most]{tcolorbox}
\usepackage{clipboard}

\newcommand{\tboxbegin}[1] {
    \begin{tcolorbox}[colback=dred!5,colframe=dred!80!black,title=\textbf{\emph{KEY TAKEAWAY {#1}}}]
}
\newcommand{\tboxend} {
	\end{tcolorbox}
}

\newcounter{ttaskno}
\DeclareRobustCommand{\ttask}[1]{%
   \refstepcounter{ttaskno}%
   \thettaskno\label{#1}}

\begin{document}


\bstctlcite{IEEEexample:BSTcontrol}
\title{
\resizebox{\linewidth}{!}{Benchmarking Memory-Centric Computing Systems:} 
\resizebox{\linewidth}{!}{Analysis of Real Processing-in-Memory Hardware}
}

\author{\IEEEauthorblockN{Juan Gómez-Luna}\IEEEauthorblockA{\textit{ETH Zürich}}
\and
\IEEEauthorblockN{Izzat El Hajj}\IEEEauthorblockA{\textit{American University}}\IEEEauthorblockA{\textit{of Beirut}}
\and
\IEEEauthorblockN{Ivan Fernandez}\IEEEauthorblockA{\textit{University}}\IEEEauthorblockA{\textit{of Malaga}}
\and
\IEEEauthorblockN{Christina Giannoula}\IEEEauthorblockA{\textit{National Technical}}\IEEEauthorblockA{\textit{University of Athens}}
\and
\IEEEauthorblockN{Geraldo F. Oliveira}\IEEEauthorblockA{\textit{ETH Zürich}}
\and
\IEEEauthorblockN{Onur Mutlu}\IEEEauthorblockA{\textit{ETH Zürich}}
}

\maketitle

\begin{abstract}

Many modern workloads such as neural network inference and graph processing are fundamentally memory-bound.
For such workloads, data movement between memory and CPU cores imposes a significant overhead in terms of both latency and energy. 
A major reason is that 
this communication happens through a narrow bus with high latency and limited bandwidth, and the low data reuse in memory-bound workloads is insufficient to amortize the cost of \om{memory} access.
Fundamentally addressing this \emph{data movement bottleneck} requires a paradigm where the memory system assumes an active role in computing by integrating processing capabilities.
This paradigm is known as \emph{processing-in-memory} (\emph{PIM}).

Recent research explores different forms of PIM architectures, motivated by the emergence of new 
technologies that integrate memory with a logic layer, where processing elements can be easily placed.
Past works evaluate these architectures in simulation or, at best, with simplified hardware prototypes.
In contrast, the UPMEM company has designed and manufactured the first publicly-available real-world PIM architecture. 
The UPMEM PIM architecture combines traditional DRAM memory arrays with general-purpose in-order cores, called \emph{DRAM Processing Units} (\emph{DPUs}), integrated in the same chip.

This paper presents key takeaways from the first comprehensive analysis~\cite{gomezluna2021benchmarking} of the first publicly-available real-world PIM architecture. First, we introduce our experimental characterization of the UPMEM PIM architecture using microbenchmarks, and present \emph{PrIM} (\emph{\underline{Pr}ocessing-\underline{I}n-\underline{M}emory benchmarks}), a benchmark suite of 16 workloads from different application domains (e.g., dense/sparse linear algebra, databases, data analytics, graph processing, neural networks, bioinformatics, image processing), which we identify as memory-bound.
Second, we provide four key takeaways about the UPMEM PIM architecture, which stem from our study of the performance and scaling characteristics of PrIM benchmarks on the UPMEM PIM architecture, and their performance and energy consumption comparison to their state-of-the-art CPU and GPU counterparts. 
More insights about suitability of different workloads to the PIM system, programming recommendations for software designers, and suggestions and hints for hardware and architecture designers of future PIM systems are available in~\cite{gomezluna2021benchmarking}.
\end{abstract}


\vspace{-3mm}

\thispagestyle{fancy}
\renewcommand{\headrulewidth}{0pt}

\section{Introduction}

In modern computing systems, a large fraction of the execution time and energy consumption of modern data-intensive workloads is spent moving data between memory and processor cores. 
This \emph{data movement bottleneck}~\cite{mutlu2019,mutlu2020modern,ghoseibm2019,mutlu2019enabling,mutlu2021intelligent} stems from the fact that, for decades, the performance of processor cores has been increasing at a faster rate than the memory performance.
The gap between an arithmetic operation and a memory access in terms of latency and energy keeps widening and the memory access is becoming increasingly more expensive.
As a result, recent experimental studies report that data movement accounts for 62\%~\cite{boroumand.asplos18} (reported in 2018), 40\%~\cite{pandiyan.iiswc2014} (reported in 2014), and 35\%~\cite{kestor.iiswc2013} (reported in 2013) of the total system energy in various consumer, scientific, and mobile applications, respectively. 

One promising way to alleviate the data movement bottleneck is \emph{processing-in-memory} (\emph{PIM}), which equips memory chips with processing capabilities~\cite{mutlu2019,mutlu2020modern,ghoseibm2019,mutlu2019enabling,mutlu2021intelligent}. 
Although this paradigm has been explored for more than 50 years~\cite{Kautz1969,stone1970logic}, limitations in memory technology prevented commercial hardware from successfully materializing. 
In recent years, the emergence of new memory innovations (e.g., 3D-stacked memories~\cite{hmc.spec.2.0, jedec.hbm.spec,lee.taco16,ghose2019demystifying,ramulator,ahn.tesseract.isca15,gokhale2015hmc}) and memory technologies (e.g., non-volatile memories~\cite{lee-isca2009, kultursay.ispass13, strukov.nature08, wong.procieee12,lee.cacm10,qureshi.isca09,zhou.isca09,lee.ieeemicro10,wong.procieee10,yoon-taco2014,yoon2012row,girard2020survey}), which aim at solving difficulties in DRAM scaling (i.e., challenges in increasing density and performance while maintaining reliability, latency and energy consumption)~\cite{kang.memoryforum14,liu.isca13,mutlu.imw13,kim-isca2014,mutlu2017rowhammer,ghose2018vampire,mutlu.superfri15,kim2020revisiting,mutlu2020retrospective,frigo2020trr,kim2018solar,raidr,mutlu2015main, mandelman.ibmjrd02, lee-isca2009,cojocar2020susceptible,yauglikcci2021blockhammer,patel2017reaper,khan.sigmetrics14,khan.dsn16,khan.micro17,lee.hpca15,lee.sigmetrics17,chang.sigmetrics17,chang.sigmetrics16,chang.hpca14,meza.dsn15,david2011memdvfs,deng2011memscale,hong2010memory,kanev.isca15,qureshi.dsn15,orosa2021rowhammer,hasan2021rowhammer}, have sparked many efforts to redesign the memory subsystem while integrating processing capabilities. 
There are two main trends among these efforts. 
\emph{Processing near memory (\emph{PNM})} integrates processing elements (e.g., functional units, accelerators, simple processing cores, reconfigurable logic) inside the logic layer of 3D-stacked memories~\cite{giannoula2021.SLS,syncron,cali2020genasm,alser2020accelerating,kim.bmc18,ahn.pei.isca15,ahn.tesseract.isca15,boroumand.asplos18,boroumand2019conda,boroumand2016pim,singh2019napel,hsieh.isca16,kim.isca16,kim.sc17,liu-spaa17,nai2017graphpim,pattnaik.pact16,pugsley2014ndc,zhang.hpdc14,DBLP:conf/isca/AkinFH15,gao2017tetris,drumond2017mondrian,dai2018graphh,zhang2018graphp,huang2020heterogeneous,zhuo2019graphq,boroumand2021polynesia,lloyd2015memory,lloyd2018dse,gokhale2015rearr,nair2015active,jacob2016compiling,sura2015data,balasubramonian2014near,boroumand2021mitigating,oliveira2021.SLS,deoliveira2021,deoliveira2021ieee,boroumand2021google}, at the memory controller~\cite{hashemi.isca16,cont-runahead}, on the \om{DRAM modules}~\cite{asghari-moghaddam.micro16,alves2015opportunities,medal2019}, or in the same package as the \om{processor} connected via silicon interposers~\cite{fernandez2020natsa,singh2020nero,singh2021fpga}. 
\emph{Processing using memory (\emph{PUM})} exploits the existing memory architecture and the operational principles of the memory cells and circuitry to perform \om{computation inside a} memory chip at low cost. 
Prior works propose PUM mechanisms using SRAM~\cite{aga.hpca17,eckert2018neural,fujiki2019duality,kang.icassp14}, DRAM~\cite{seshadri2020indram,Seshadri:2015:ANDOR,seshadri.micro17,seshadri2018rowclone,seshadri2013rowclone,kim.hpca18,kim.hpca19,gao2020computedram,chang.hpca16,li.micro17,deng.dac2018,hajinazarsimdram,rezaei2020nom,wang2020figaro,ali2019memory,ferreira2021pluto,olgun2021quactrng,olgun2021.SLS,hajinazar2021.SLS,rezaei2020nom,kim.hpca18,orosa2021codic}, 
PCM~\cite{li.dac16}, 
MRAM~\cite{angizi2018pima,angizi2018cmp,angizi2019dna}, 
or RRAM/memristive~\cite{levy.microelec14,kvatinsky.tcasii14,shafiee2016isaac,kvatinsky.iccd11,kvatinsky.tvlsi14,gaillardon2016plim,bhattacharjee2017revamp,hamdioui2015memristor,xie2015fast,hamdioui2017myth,yu2018memristive,puma-asplos2019, ankit2020panther,chi2016prime,song2018graphr,zheng2016tcam,xi2020memory} memories. 

\begin{figure*}[t]
    \centering
    \includegraphics[width=0.95\linewidth]{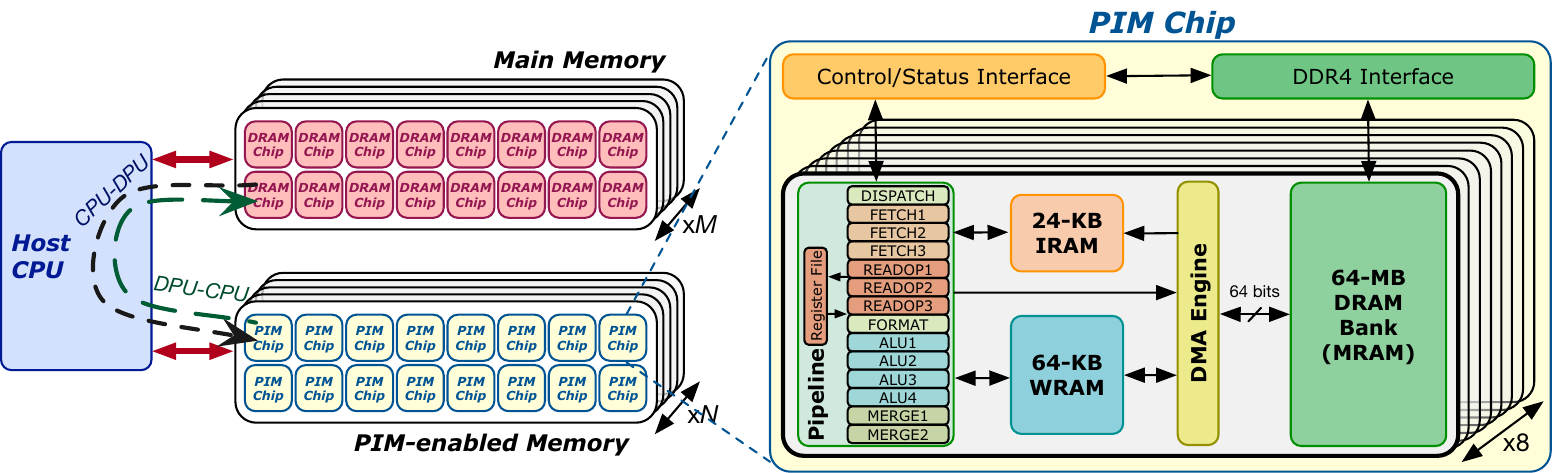}
    \caption{UPMEM-based PIM system with a host CPU, standard main memory, and PIM-enabled memory (left), and internal components of a UPMEM PIM chip (right)~\cite{upmem2018, devaux2019}.}
    \label{fig:scheme}
    \vspace{-1mm}
\end{figure*}

\begin{table*}[b]
\begin{center}
\caption{PrIM benchmarks~\cite{gomezluna2021repo}.}
\label{tab:benchmarks}
\resizebox{0.98\linewidth}{!}{
\begin{tabular}{|l|l|c||c|c|c|c|c|c|c|}
    \hline
    \multirow{2}{*}{\textbf{Domain}} & \multirow{2}{*}{\textbf{Benchmark}} & \multirow{2}{*}{\textbf{Short name}} & \multicolumn{3}{c|}{\textbf{Memory access pattern}} & \multicolumn{2}{c|}{\textbf{Computation pattern}} & \multicolumn{2}{c|}{\textbf{Communication/synchronization}}  \\
    \cline{4-10}
     & & & \textbf{Sequential} & {\textbf{Strided}} & \textbf{Random} & \textbf{Operations} & \textbf{Datatype} & \textbf{Intra-DPU} & \textbf{Inter-DPU}  \\
    \hline
    \hline
    \multirow{2}{*}{Dense linear algebra}
      & Vector Addition & VA & Yes & & & add & int32\_t &  &  \\
    \cline{2-10}
      & Matrix-Vector Multiply & GEMV & Yes & & & add, mul & uint32\_t &  &  \\
    \hline
    \multirow{1}{*}{Sparse linear algebra}
      & Sparse Matrix-Vector Multiply & SpMV & Yes & & Yes & add, mul & float &  &  \\
    \hline
    \multirow{2}{*}{Databases}
      & Select & SEL & Yes & & & add, compare & int64\_t & handshake, barrier & Yes \\
    \cline{2-10}
      & Unique & UNI & Yes & & & add, compare & int64\_t & handshake, barrier & Yes \\
    \hline
    \multirow{2}{*}{Data analytics}
      & Binary Search & BS & {Yes} & & Yes & compare & int{64}\_t &  &  \\
    \cline{2-10}
      & Time Series Analysis & TS & Yes & & & add, sub, mul, div & int32\_t &  &  \\
    \hline
    \multirow{1}{*}{Graph processing}
      & Breadth-First Search & BFS & Yes & & Yes & bitwise logic & uint64\_t & barrier, mutex & Yes \\
    \hline
    \multirow{1}{*}{Neural networks}
      & Multilayer Perceptron & MLP & Yes & & & add, mul, compare & int32\_t &  &  \\
    \hline
    \multirow{1}{*}{Bioinformatics}
      & Needleman-Wunsch & NW & Yes & Yes & & add, sub, compare & int32\_t & barrier & Yes \\
    \hline
    \multirow{2}{*}{Image processing}
      & Image histogram (short) & {HST-S} & Yes & & Yes & add & {uint32\_t} & barrier & Yes \\
    \cline{2-10}
      & Image histogram (long) & {HST-L} & Yes & & Yes & add & {uint32\_t} & barrier, mutex & Yes \\
    \hline
    \multirow{4}{*}{Parallel primitives}
      & Reduction & RED & Yes & Yes & & add & int64\_t & barrier & Yes \\
    \cline{2-10}
      & Prefix sum (scan-scan-add) & SCAN-SSA & Yes & & & add & int64\_t & handshake, barrier & Yes \\
    \cline{2-10}
      & Prefix sum (reduce-scan-scan) & SCAN-RSS & Yes & & & add & int64\_t & handshake, barrier & Yes \\
    \cline{2-10}
      & Matrix transposition & TRNS & Yes & & Yes & add, sub, mul & int64\_t & mutex & \\
    \hline
\end{tabular}
}
\end{center}
\vspace{-4mm}
\end{table*}

The UPMEM company has designed and fabricated the first commercially-available PIM architecture. 
The UPMEM PIM architecture~\cite{upmem2018, devaux2019, gomezluna2021benchmarking,gomezluna2021.SLS,gomezluna2020.lecture} combines traditional DRAM memory arrays with general-purpose in-order cores, called \emph{DRAM Processing Units} (\emph{DPUs}), integrated in the same DRAM chip. 
UPMEM PIM chips are mounted on DDR4 memory modules that coexist with regular DRAM modules (i.e., the main memory) attached to a \emph{host} CPU.
Figure~\ref{fig:scheme} (left) depicts a UPMEM-based PIM system with (1) a host CPU, (2) main memory (DRAM memory modules), and (3) PIM-enabled memory (UPMEM modules). PIM-enabled memory can reside on one or more memory channels.

Inside each UPMEM PIM chip (Figure~\ref{fig:scheme} (right)), there are 8 DPUs. Each DPU 
has exclusive access to (1) a 64-MB DRAM bank, called \emph{Main RAM} (\emph{MRAM}), (2) a 24-KB instruction memory, and (3) a 64-KB scratchpad memory, called \emph{Working RAM} (\emph{WRAM}). 
The MRAM banks are accessible by the host CPU for \emph{copying} input data (from main memory to MRAM) and \emph{retrieving} results (from MRAM to main memory). 
These data transfers can be performed in parallel (i.e., concurrently across multiple MRAM banks), if the size of the buffers transferred from/to all MRAM banks is the same. Otherwise, the data transfers happen serially. 
There is no support for direct communication between DPUs. All inter-DPU communication takes place through the host CPU by \emph{retrieving} results and \emph{copying} data.

Rigorously understanding the UPMEM PIM architecture, the first publicly-available PIM architecture, and its suitability to various workloads can provide valuable insights to programmers, users and architects of this architecture as well as of future PIM systems. 
To this end, our work~\cite{gomezluna2021benchmarking,gomezluna2021.SLS} provides the first comprehensive analysis of the first publicly-available real-world PIM architecture.
We make two key contributions. 
First, we conduct an experimental characterization of the UPMEM-based PIM system using microbenchmarks to assess various architecture limits such as compute throughput and memory bandwidth, yielding new insights.
Second, we present \emph{PrIM} (\emph{\underline{Pr}ocessing-\underline{I}n-\underline{M}emory benchmarks}), an \om{open-source} benchmark suite~\cite{gomezluna2021repo} of 16 workloads from different application domains (e.g., neural networks, databases, graph processing, bioinformatics), which we identify as \emph{memory-bound} workloads using the roofline model~\cite{roofline} (i.e., these workloads' performance in conventional processor-centric architectures is limited by memory access). 
Table~\ref{tab:benchmarks} shows a summary of PrIM benchmarks, including workload characteristics (memory access pattern, computation pattern, communication/synchronization needs) that demonstrate the diversity of the benchmarks.

Our comprehensive analysis~\cite{gomezluna2021benchmarking,gomezluna2021.SLS} evaluates the performance and scaling characteristics of PrIM benchmarks on the UPMEM PIM architecture, and compares their performance and energy consumption to their CPU and GPU counterparts.
Our extensive evaluation conducted on two real UPMEM-based PIM systems with 640 and 2,556 DPUs provides new insights about suitability of different workloads to the PIM system, programming recommendations for software designers, and suggestions and hints for hardware and architecture designers of future PIM systems.

In this paper, we provide four key takeaways that represent the main insights and conclusions of our work~\cite{gomezluna2021benchmarking,gomezluna2021.SLS}. 
For more information about our thorough PIM architecture characterization, methodology, results, insights, and the PrIM benchmark suite, we refer the reader to the full version of the paper~\cite{gomezluna2021benchmarking,gomezluna2021.SLS}. We hope that our study can guide programmers on how to optimize software for real PIM systems and enlighten designers about how to improve the architecture and hardware of future PIM systems. 
Our microbenchmarks and PrIM benchmark suite are publicly available~\cite{gomezluna2021repo}.


\section{Key Takeaways}\label{sec:discussion}

We present several key empirical observations in the form of four key takeaways that \om{we distill} from our experimental characterization of the UPMEM PIM architecture~\cite{gomezluna2021benchmarking}. We also provide \om{analyses of} workload suitability and good programming practices for the UPMEM PIM architecture, and suggestions for hardware and architecture designers of future PIM systems.

\begin{figure}[h]
        \centering
        \includegraphics[width=0.95\linewidth]{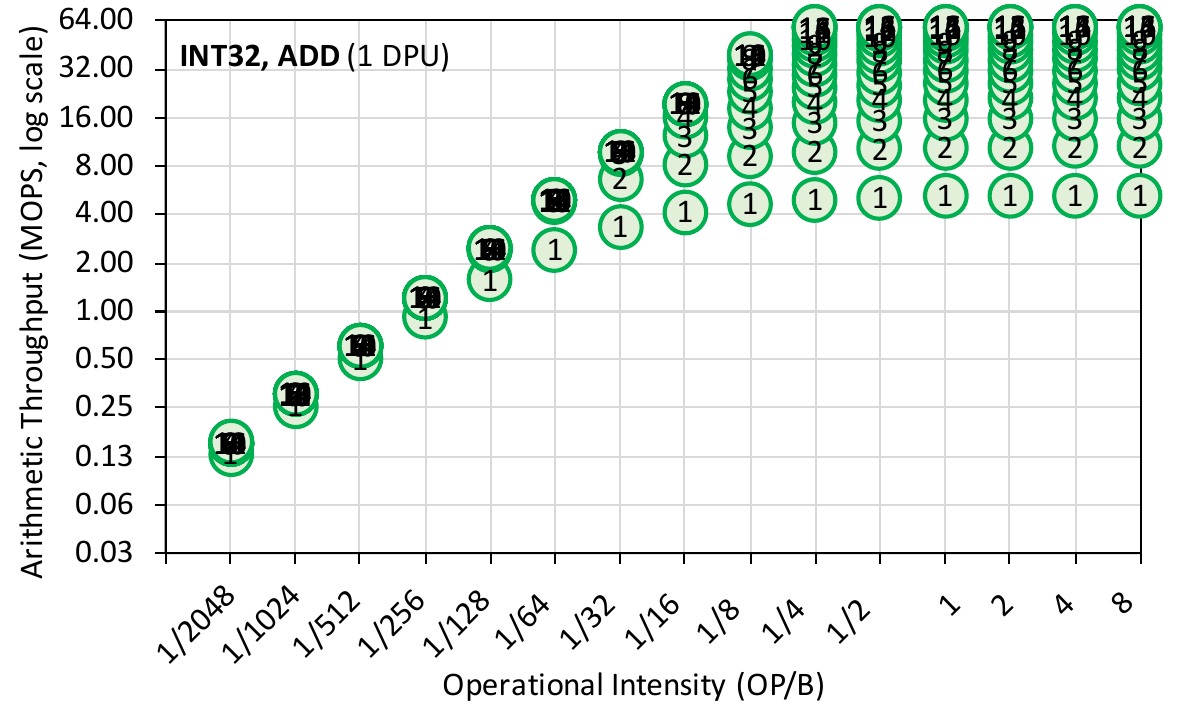}
        \caption{Arithmetic throughput versus operational intensity for 32-bit integer addition. The number inside each dot indicates the number of tasklets. 
        Both x- and y-axes are log scale.}
        \label{fig:ai-dpu}
        \vspace{-2mm}
\end{figure}

\noindent\textbf{Key Takeaway \#1}. 
\textbf{The UPMEM PIM architecture is fundamentally compute bound}.
Our microbenchmark-based analysis shows that workloads with more complex operations than integer addition fully utilize the instruction pipeline before they can potentially saturate the memory bandwidth.
As Figure~\ref{fig:ai-dpu} shows, even workloads with as simple operations as integer addition saturate the compute throughput with an operational intensity as low as 0.25 operations/byte (1 addition per integer accessed).

\vspace{-1mm}

This key takeaway shows that \textbf{the most suitable workloads for the UPMEM PIM architecture are memory-bound workloads}. 
From a programmer's perspective, the architecture requires a shift in how we think about computation and data access, since the relative cost of computation vs. data access in the PIM system is very different from that in the dominant processor-centric architectures of today.

\tboxbegin{\ttask{kt}}
\vspace{-1mm}
\textbf{The UPMEM PIM architecture is fundamentally compute bound}. 
As a result, \textbf{the most suitable workloads are memory-bound}.
\vspace{-1mm}
\tboxend

\begin{figure}[h]
    \centering
    \includegraphics[width=0.95\linewidth]{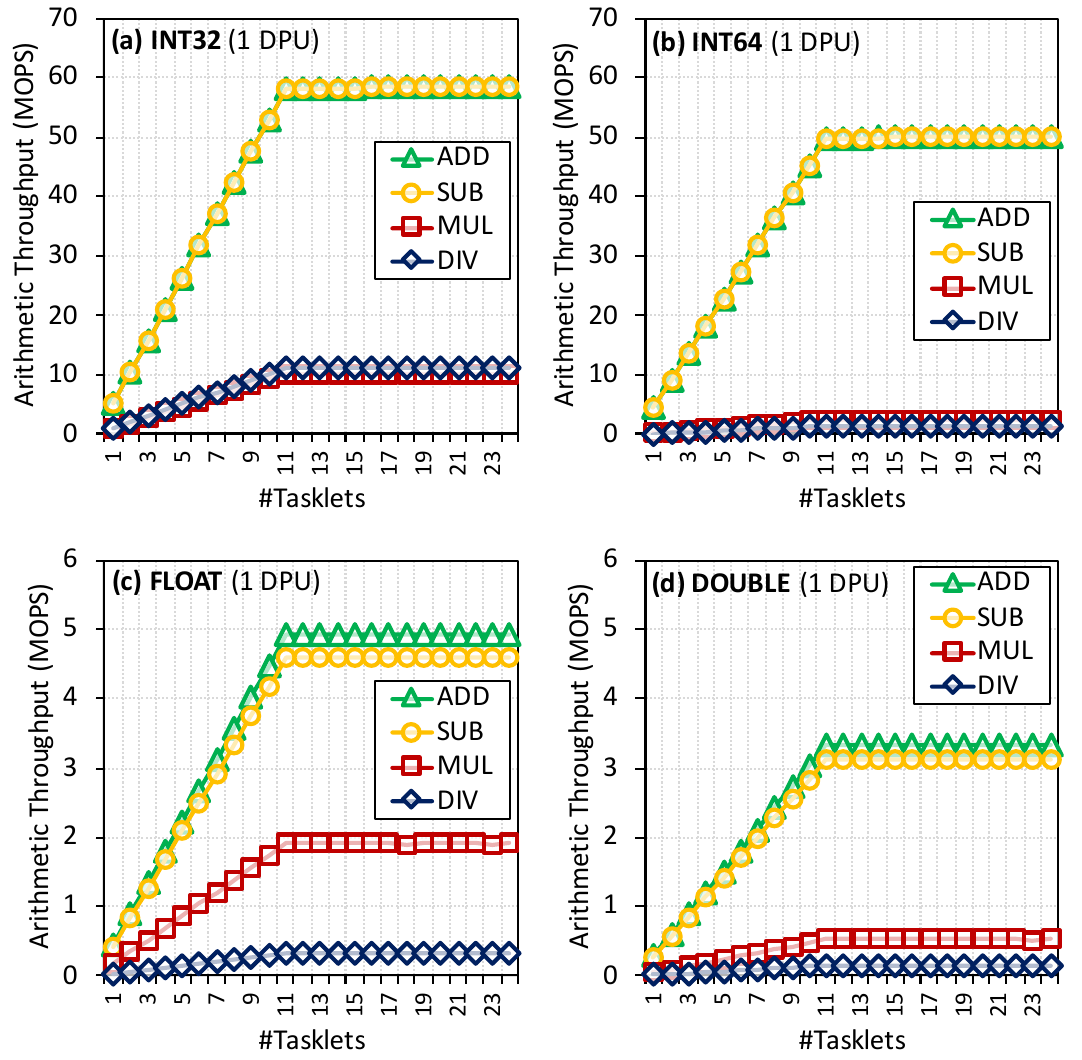}
    \vspace{-1mm}
    \caption{Throughput of arithmetic operations (ADD, SUB, MUL, DIV) on one DPU for four different data types: (a) INT32, (b) INT64, (c) FLOAT, (d) DOUBLE.}
    \label{fig:throughput-dpu}
    \vspace{-2mm}
\end{figure}

\noindent\textbf{Key Takeaway \#2}. 
\textbf{The workloads most well-suited for the UPMEM PIM architecture are those with simple or no arithmetic operations}. This is because DPUs include native support for \emph{only} integer addition/subtraction and bitwise operations. More complex integer (e.g., multiplication, division) and floating point operations are implemented using software library routines.
As Figure~\ref{fig:throughput-dpu} shows, the arithmetic throughput of more complex integer operations and floating point operations are an order of magnitude lower than that of simple addition and subtraction.

\vspace{-2mm}

\begin{figure}[h]
        \includegraphics[width=\linewidth]{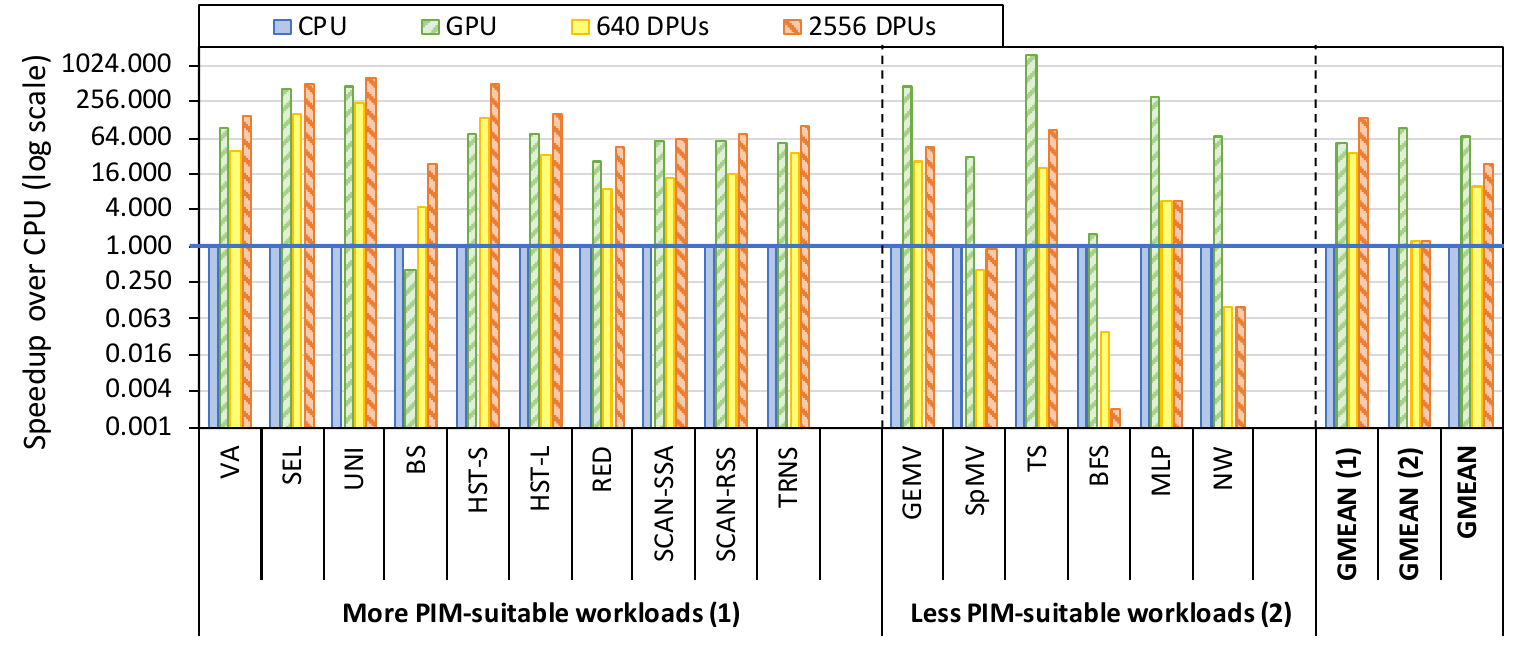}
        \caption{Performance comparison between the UPMEM-based PIM systems with 640 and 2,556 DPUs, a Titan V GPU, and an Intel Xeon E3-1240 CPU. Results are normalized to the CPU performance (y-axis is log scale). 
        There are two groups of benchmarks: (1) benchmarks that are more suitable to the UPMEM PIM architecture, and (2) benchmarks that are less suitable to the UPMEM PIM architecture.} \label{fig:comparison-perf}
\end{figure}

Figure~\ref{fig:comparison-perf} shows the speedup of the UPMEM-based PIM systems with 640 and 2,556 DPUs and a state-of-the-art Titan V GPU over a state-of-the-art Intel Xeon CPU. 

We observe that benchmarks with little amount of computation and no use of multiplication, division, or floating point operations (10 out of 16 benchmarks) run faster (2.54$\times$ on average) on a 2,556-DPU system than on a state-of-the-art NVIDIA Titan V GPU. 
These observations show that \textbf{the workloads most well-suited for the UPMEM PIM architecture are those with no arithmetic operations or simple operations (e.g., bitwise operations and integer addition/subtraction)}.
Based on this key takeaway, we recommend devising much more efficient software library routines or, more importantly, specialized and fast in-memory hardware for complex operations in future PIM architecture generations to improve the general-purpose performance of PIM systems.

\tboxbegin{\ttask{kt}}
\vspace{-1mm}
\textbf{The most well-suited workloads for the UPMEM PIM architecture use no arithmetic operations or use only simple operations (e.g., bitwise operations and integer addition/subtraction).}
\vspace{-1mm}
\tboxend

\noindent\textbf{Key Takeaway \#3}. 
\textbf{The workloads most well-suited for the UPMEM PIM architecture are those with little global communication}, because there is no direct communication channel among DPUs (see Figure~\ref{fig:scheme}). 
As a result, there is a huge disparity in performance scalability of benchmarks that do \emph{not} require inter-DPU communication and benchmarks that do (especially if parallel transfers across MRAM banks cannot be used). 
This key takeaway shows that \textbf{the workloads most well-suited for the UPMEM PIM architecture are those with little or no inter-DPU communication}.
Based on this takeaway, we recommend that the hardware architecture and the software stack be enhanced with support for inter-DPU communication (e.g., by leveraging new in-DRAM data copy techniques~\cite{seshadri2013rowclone, wang2020figaro, rezaei2020nom, chang.hpca16} and providing better connectivity inside DRAM~\cite{chang.hpca16, rezaei2020nom}).

\tboxbegin{\ttask{kt}}
\textbf{The most well-suited workloads for the UPMEM PIM architecture require little or no communication across DRAM Processing Units (inter-DPU communication)}.
\tboxend

\noindent\textbf{Summary}. 
We find that the workloads most suitable for the UPMEM PIM architecture in its current form are (1) memory-bound workloads with (2) simple or no arithmetic operations and (3) little or no inter-DPU communication. 

\noindent\textbf{Key Takeaway \#4}. 
We observe that the existing UPMEM-based PIM systems greatly improve energy efficiency and performance over state-of-the-art CPU and GPU systems across many workloads we examine. Figure~\ref{fig:comparison-perf} shows that the 2,556-DPU and the 640-DPU systems are 23.2$\times$ and 10.1$\times$ faster, respectively, than a state-of-the-art Intel Xeon CPU, averaged across the entire set of 16 PrIM benchmarks. 
We also observe that the 640-DPU system is 1.64$\times$ more energy efficient than the CPU, averaged across the entire set of 16 PrIM benchmarks, and 5.23$\times$ more energy efficient for 12 of the PrIM benchmarks. 

The 2,556-DPU system is faster (on average by 2.54$\times$) than the state-of-the-art GPU in 10 out of 16 PrIM benchmarks, which have three key characteristics that define a workload's PIM suitability: (1) streaming memory accesses, (2) little or no inter-DPU communication, and (3) little or no use of multiplication, division, or floating point operations. 

We expect that the 2,556-DPU system will {provide even higher} performance and energy benefits, and that future PIM systems will be even better (especially after implementing our recommendations for future PIM hardware~\cite{gomezluna2021benchmarking}).
If the architecture is improved based on our recommendations under Key Takeaways 1-3 \om{and in~\cite{gomezluna2021benchmarking}}, we believe future PIM systems will be even more attractive, leading to much higher performance and energy benefits {versus state-of-the-art CPUs and GPUs} over potentially all workloads.

\tboxbegin{\ttask{kt}}
\begin{itemize}[wide, labelsep=0.5em]
\item UPMEM-based PIM systems \textbf{outperform state-of-the-art CPUs in terms of performance} (by 23.2$\times$ on 2,556 DPUs for 16 PrIM benchmarks) \textbf{and energy efficiency} (by 5.23$\times$ on 640 DPUs for 12 PrIM benchmarks). 
\item UPMEM-based PIM systems \textbf{outperform state-of-the-art GPUs on a majority of PrIM benchmarks} (by 2.54$\times$ on 2,556 DPUs for 10 PrIM benchmarks), and the outlook is even more positive for future PIM systems.
\item {UPMEM-based PIM systems are \textbf{more energy-efficient than state-of-the-art CPUs and GPUs on workloads that they provide performance improvements} over the CPUs and the GPUs.}
\end{itemize}
\tboxend

\section{Summary \& Conclusion}
This \om{invited} short paper \om{summarizes} the first comprehensive characterization and analysis of a real commercial PIM architecture~\cite{gomezluna2021benchmarking,gomezluna2021.SLS}. 
Through this analysis, we develop a rigorous, thorough understanding of the UPMEM PIM architecture, the first publicly-available PIM architecture, and its suitability to various types of workloads.

First, we conduct a characterization of the UPMEM-based PIM system using microbenchmarks to assess various architecture limits such as compute throughput and memory bandwidth, yielding new insights. 
Second, we present PrIM, an \om{open-source} benchmark suite~\cite{gomezluna2021repo} of 16 memory-bound workloads from different application domains (e.g., dense/sparse linear algebra, databases, data analytics, graph processing, neural networks, bioinformatics, image processing).

Our extensive evaluation of PrIM benchmarks conducted on two real systems with UPMEM memory modules provides new insights about suitability of different workloads to the PIM system, programming recommendations for software designers, and suggestions and hints for hardware and architecture designers of future PIM systems. 
We compare the performance and energy consumption of the UPMEM-based PIM systems for PrIM benchmarks to those of a state-of-the-art CPU and a state-of-the-art GPU, and identify key workload characteristics that can successfully leverage the strengths of a real PIM system over conventional processor-centric architectures, \om{leading to significant performance and energy improvements}.

We believe and hope that our work will provide valuable insights to programmers, users and architects of this PIM architecture as well as of future PIM systems, and will represent an enabling milestone in the development of \om{fundamentally-efficient} memory-centric computing systems.

\section*{Acknowledgment}
We thank UPMEM's Fabrice Devaux, Rémy Cimadomo, Romaric Jodin, and Vincent Palatin for their valuable support.
We acknowledge the support of SAFARI Research Group's industrial partners, especially ASML, Facebook, Google, Huawei, Intel, Microsoft, VMware, and the Semiconductor Research Corporation.
This research was partially supported by the ETH Future Computing Laboratory. 
Izzat El Hajj acknowledges the support of the University Research Board of the American University of Beirut (URB-AUB-103951-25960).
\om{This paper provides a short summary of our larger paper on arxiv.org~\cite{gomezluna2021benchmarking}, which has comprehensive descriptions and extensive analyses.}

\balance

\bibliographystyle{IEEEtran}
\bibliography{references}

\end{document}